\documentclass[letter,twocolumn]{jpsj3}%
\usepackage{amsmath}
\usepackage{txfonts}
\usepackage[usenames]{color}
\usepackage{color}
\usepackage{amsfonts}
\usepackage{amssymb}
\usepackage{graphicx}%
\providecommand{\U}[1]{\protect\rule{.1in}{.1in}}
\providecommand{\U}[1]{\protect\rule{.1in}{.1in}}

\inst{
$^1$Department of Applied Physics, Faculty of Science,
Tokyo University of Science, Shinjuku, Tokyo 162-8601, Japan\\
$^2$Division of Natural and Environmental Sciences, The Open University of Japan,
Chiba, 261-8586, Japan\\
$^3$Department of Physics, University of Tokyo,
Hongo 7-3-1, Bunkyo, Tokyo 113-0033, Japan
}
\abst{
A possible theoretical scheme is proposed for the unified
description of charge excitations with $10^{10}$ orders of magnitude
difference in energy [from optical($\sim$eV) down to dielectric anomalies
(10KHz$\sim 10^{-10}$eV)] based on a simple and single model.
}
\begin{document}

\title{Energy Landscape of Charge Excitations in Boundary Region between Dimer-Mott
and Charge Order State in Molecular Solids}
\author{Hidetoshi Fukuyama$^{1}$, Jun-ichiro Kishine$^{2}$, and Masao Ogata$^{3}$}
\maketitle

It has been established that the variety of ground states of
strongly-correlated quarter-filled molecular solids can be systematically
understood by noting dimer-Mott (DM) and charge ordering (CO) states as
limiting cases controlled by the degree of dimerization.\cite{SeoReview}
Recent experiments have disclosed remarkable and unexpected features of
existence of charge excitations in such DM type insulators based on BEDT-TTF
(ET) molecules: i.e., the existence of anomalous dielectric responses in
$\kappa$-ET$_{2}$Cu$_{2}$(CN)$_{3}$ for frequency $\omega\sim10$ KHz and in
the temperature range $T<50$K,\cite{AbdelJawad} while in $\beta^{\prime}%
$-ET$_{2}$ICl$_{2}$ for $\omega\sim100$ KHz and $T<100$K.\cite{Iguchi} These
experimental findings have disclosed the fact that charge fluctuations are
possible even in apparent DM states and then these DM states are not in the
limit of strong dimerization. These experimental findings have stimulated
search for the exploration of new possibilities near the DM-CO boundary.

Theoretically, there are some
studies\cite{AbdelJawad,Ishihara,Hotta2010,Hotta2012} addressing to these
intriguing properties of charge excitations near the DM-CO boundaries based on
the tight-binding model in two-dimension with both dimerization and Coulomb
interaction on site and inter site. We note, however, that the energy scale of
the observed phenomena as dielectric anomalies is very small, e.g.,
$\hbar\omega\sim10^{-10}-10^{-9}$ eV for $\omega\sim10$K$-$100KHz, and then
almost 10 orders of magnitude smaller than those in optical and infrared
absorption with energy scale of the order of 1 eV, only the latter of which
can be studied by the tight-binding model in a conventional way. In this paper
we propose theoretically a possible scenario linking between such large energy
differences based on a model Hamiltonian. Although the present theoretical
studies are motivated by experiments on two-dimensional molecular solids, we
adopt one-dimensional model because of mathematical transparency, and then
comparison of the theoretical results with experimental findings is taken not
truly quantitative but qualitative.

Our starting microscopic Hamiltonian is given as follows.%
\begin{align}
\mathcal{H}  &  =-\sum_{j,\sigma}[t+(-1)^{j}t_{d}](c_{j,\sigma}^{\dagger
}c_{j+1,\sigma}+\mathrm{h.c.})\nonumber\\
&  +U\sum_{j}n_{j,\uparrow}n_{j,\downarrow}+V\sum_{j}n_{j}n_{j+1},
\end{align}
where quarter-filling of the band is assumed, i.e., average electron number
per site being 1/2. In the presence of finite dimerization, $t_{d}$, the
original band is folded and becomes half-filled in the reduced Brillouin zone,
and hence $t_{d}$ bridges between 1/4-and 1/2-filling smoothly. For
theoretical studies of low energy excitations (compared with the Fermi energy
$\varepsilon_{\mathrm{F}}$) of one-dimensional systems such as the one
represented by eq.~(1), the phase Hamiltonian approach is very transparent and
effective.\cite{Suzumura,TakaFuku} With finite $t_{d}$, eq.~(1) can be
transformed into the following\cite{Tsuchiizu} in terms of phase variables,
$\theta(x)$, describing charge degrees of freedom where $[\theta
(x),P(x^{\prime})]=i\delta(x-x^{\prime})$,%
\begin{align}
\mathcal{H}  &  =\int dx\biggl[A_{\rho}(\partial_{x}\theta_{+})^{2}+C_{\rho
}P(x)^{2}\nonumber\\
&  -(g_{1/2}\sin2\theta+g_{1/4}\cos4\theta)/{2\pi^{2}\alpha^{2}}\biggr],
\label{PhaseH}%
\end{align}
where $A_{\rho}=\left(  \hbar v_{F}/4\pi\right)  \left[  1+(U+4V)a/\pi\hbar
v_{F}\right]  ,$ $C_{\rho}=\pi\hbar v_{F},$ $g_{1/2}=BUa,$ $g_{1/4}=a^{3}%
U^{2}(4V-U)/4\hbar^{2}v_{\mathrm{F}}^{2}$ with $\hbar v_{\mathrm{F}}%
=2ta\sin(k_{\mathrm{F}}a)=2ta\sin(\pi/4)=\sqrt{2}ta$ and $\alpha=a/\pi
$,\cite{Nakano} are respectively Fermi velocity and
cut-off parameter, where $a$ is the lattice constant, and{\ $B=2t_{d}/t$}.
Equation (\ref{PhaseH}) can further be rewritten as follows in terms of
$\varphi=\theta-\pi/4$.
\begin{equation}
\mathcal{H}=\varepsilon_{\mathrm{F}}\int dy\left[  A(\nabla\varphi)^{2}%
+P^{2}+\mathcal{V}(\varphi)\right]  , \label{PhaseH2}%
\end{equation}
where%
\begin{equation}
A=\frac{1}{4\pi^{2}}(1+u+4v), \label{A}%
\end{equation}
with $y=x/a,$ $u=U/\varepsilon_{\mathrm{F}},$ $v=V/\varepsilon_{\mathrm{F}},$
and $\varepsilon_{\mathrm{F}}=\pi\hbar v_{\mathrm{F}}/a$ where $\varepsilon
_{\mathrm{F}}$ is the Fermi energy. The potential $\mathcal{V}(\varphi)$ is
defined by
\begin{equation}
\mathcal{V}(\varphi)=-g_{D}\cos2\varphi+g_{C}\cos4\varphi, \label{potential}%
\end{equation}
where
\begin{align}
g_{D}  &  =\frac{g_{1/2}}{2a\varepsilon_{\mathrm{F}}}=\frac{t_{d}}%
{t}u,\label{gD}\\
g_{C}  &  =\frac{g_{1/4}}{2a\varepsilon_{\mathrm{F}}}=\frac{\pi^{2}}{8}%
u^{2}\left(  4v-u\right)  . \label{gC}%
\end{align}
In $\mathcal{V}(\varphi)$, eq.~(\ref{potential}), $\varphi=0$ corresponds to
DM state, while $\varphi=-\pi/4$ or $\pi/4$ to CO states, respectively. It is
seen that $\mathcal{V}(\varphi)$ takes minimum at $\varphi=0$, pure dimer-Mott
state, as far as $g_{D}>4g_{C}$, while this state becomes unstable once
$g_{D}<4g_{C}$. Actually $\mathcal{V}(\varphi)$ around pure DM state
($\varphi\sim0$) is given by
\begin{align}
\mathcal{V}(\varphi)  &  =(g_{C}-g_{D})-\frac{K_{2}}{2}\varphi^{2}+\frac
{K_{4}}{4}\varphi^{4}+\mathcal{O}(\varphi^{6})\nonumber\\
&  =(g_{C}-g_{D})+\frac{K_{4}}{4}\left(  \varphi^{2}-\varphi_{0}^{2}\right)
^{2}-\frac{K_{2}^{2}}{4K_{4}}+\mathcal{O}(\varphi^{6}), \label{PhaseH3}%
\end{align}
with $K_{2}=4(4g_{C}-g_{D})$ and $K_{4}=\frac{8}{3}(16g_{C}-g_{D})$. {In
Fig.~1, }$\varphi${-dependences of }$\mathcal{V}${$(\varphi)$ are shown with
choices of $g_{D}=5$ and $g_{C}=1$~(a), and $g_{D}=3$ and $g_{C}=1$~(b), where
the DM state and the CO state are the ground state, respectively.}

In the following, we assume $K_{2}>0$, in which case minima of $\mathcal{V}%
(\varphi)$ locate at $\varphi=\pm\varphi_{0}$ with $\varphi_{0}=(K_{2}%
/K_{4})^{1/2}$ [{Fig.~1 (b)}]. We will study the case where $g_{D}$ is
appreciably larger than $g_{C}$ but not so large so that $g_{D}<4g_{C}$ is
satisfied, which will be the case when DM state compete with CO state. From
now on, we use the fixed choice of parameters, $g_{D}=3$ and $g_{C}=1$.{\ }

\begin{figure}[h]
\begin{center}
\includegraphics[width=8cm]{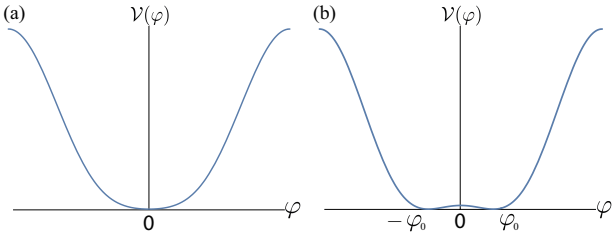} \vskip 0.5truecm
\end{center}
\caption{ {\ Potential }$\mathcal{V}${$(\varphi)$ with (a) $\left(
g_{D},g_{C}\right)  =\left(  5,1\right)  $ and (b) $\left(  g_{D}%
,g_{C}\right)  =\left(  3,1\right)  $, where the DM and CO states are the
ground states, respectively.}}%
\label{Fig:Gra}%
\end{figure}

We will show that there exist four different characteristic energy scales,
$E_{1}$, $E_{2}$, $E_{3}$, and $E_{4}$ , of charge excitations in the case as
shown in Fig.~1(b). Two energy scales ($E_{1}$, $E_{2}$) are in high energy
quantum region, which are optical absorption across the Mott gap, $E_{1}$, and
uniform phase oscillation (phason) expected to be in infrared region, $E_{2}$.
Another two energy scales ($E_{3}$, $E_{4}$) are in low energy classical
region, which are small uniform oscillation around either $\varphi=\varphi
_{0}$ or $\varphi=-\varphi_{0}$, $E_{3}$, and large amplitude spatial
modulation bridging between them as described as a domain wall or a soliton to
be associated with dielectric response, $E_{4}$%
.\cite{comment,Sasaki2004,Sasaki2005}


The largest energy scale of charge excitations, $E_{1}$, governed by
eq.~(\ref{PhaseH2}) is the one due to $g_{1/2}$, which can be estimated by
mapping eq.~(\ref{PhaseH2}) (ignoring $g_{1/4}$) to the massive Thirring model
of spinless fermions as described in \cite{Mori}. This corresponds to
excitations across the Mott gap, and\ leads to
\begin{equation}
E_{1}\sim\frac{2g_{1/2}}{\pi\alpha}=2\left(  \frac{t_{d}}{t}\right)
U=\varepsilon_{\mathrm{F}}\left(  2\frac{t_{d}}{t}\right)  u. \label{E1}%
\end{equation}

At lower energies, phasons start to be affected by the details of the
restoring forces resulting from finite $g_{C}$ in eq.~(\ref{potential}). To
assess this, $\mathcal{V}(\varphi)$ is treated as follows in a way similar to
the self-consistent harmonic approximation,%
\begin{align}
\mathcal{V}(\varphi)  &  =-g_{D}\cos2\varphi+g_{C}(2\cos^{2}2\varphi
-1)\nonumber\\
&  \sim-g_{D}\cos2\varphi+4g_{C}\langle\cos2\varphi\rangle\cos2\varphi
\nonumber\\
&  =-\tilde{g}_{D}\cos2\varphi=-\tilde{g}_{D}(1-2\varphi^{2}),
\end{align}
where
\begin{equation}
\tilde{g}_{D}=g_{D}-4g_{C}\langle\cos2\varphi\rangle=g_{D}-4g_{C}%
e^{-2\langle\varphi^{2}\rangle}.
\end{equation}
Here for $\tilde{g}_{D}$, frequency of uniform phason $\omega_{0}$ around the
pure DM state, $\varphi=0$, governed by eq.~({\ref{PhaseH2}) is given
$\omega_{0}^{2}=8\tilde{g}_{D}\varepsilon_{\mathrm{F}}^{2}/\hbar^{2}$, i.e.,}%
\begin{equation}
E_{2}=\hbar{\omega_{0}}=2\sqrt{2\tilde{g}_{D}}\varepsilon_{\mathrm{F}}.
\label{E2}%
\end{equation}
{\ The condition }${0<}${$\tilde{g}_{D}$ together with }${0<}${$K_{2}$ leads
to
\begin{equation}
4g_{C}e^{-2\langle\varphi^{2}\rangle}<g_{D}<4g_{C}. \label{SCeqGC}%
\end{equation}
Eq.~(\ref{SCeqGC}) will be easily satisfied since $\langle\varphi^{2}\rangle$
is larger than the case of maximum restoring force $\tilde{g}_{D}$, i.e.,
$g_{D}$, for which
\begin{equation}
\exp[{-2\langle\varphi^{2}\rangle}]<\exp\left(  -\frac{{1}}{\sqrt{2g_{D}}%
}\right)  =\exp\left(  -\frac{{1}}{\sqrt{6}}\right)  =0.665,
\end{equation}
for the present choice of $g_{D}=3$. }

At very low frequencies, we have to look more in detail of $\mathcal{V}%
(\varphi)$ which represents double potential minima for the case of
$g_{D}<4g_{C}$ indicating that pure dimer Mott state is unstable as seen in
Fig.~1(b). This implies the intriguing fact that, if the quantum fluctuations
are suppressed, pure dimer Mott states no longer exist, but have always small
charge disproportionation with either $\varphi=\varphi_{0}$ or $\varphi
=-\varphi_{0}$. Actually, in such low energy region the dynamics of phase
variable will be treated classically since the quantum fluctuations intrinsic
to purely one-dimensional systems as expressed by eq.~({\ref{PhaseH2}) will no
longer be the case because of the presence of three dimensionality of real
systems.}

The dynamical correlation in such situation with double potential minima in
one-dimensional systems has been studied in the context of structural phase
transition, which has clarified the existence of two different kinds of
excitations, small amplitude oscillations around potential minima
($\varphi_{0}$ or $-\varphi_{0}$, here) on one hand, $E_{3}$, and large
amplitude spatial modulation bridging between the two minima described as a
domain wall or a soliton, on the other hand. (In the structural phase
transition the former is called as phonons, while the latter central
modes\cite{Scott}). In the following we focus on the latter, $E_{4}%
$.\cite{comment,Sasaki2004,Sasaki2005}

For the present model defined by eqs.~(\ref{PhaseH2}) and (\ref{PhaseH3}), the
solution of an isolated soliton, $\varphi_{s}$, is given by
\begin{equation}
\varphi_{s}=\sqrt{\frac{K_{2}}{K_{4}}}\tanh\left[  (\bar{y}-\bar{v}_{d}\bar
{t})/\sqrt{2}\right]  ,
\end{equation}
where $\bar{y}=y/y_{0},\bar{t}=\left(  \varepsilon_{\mathrm{F}}/\hbar\right)
\sqrt{2K_{2}}t,\bar{v}_{d}=\hbar v_{d}/(y_{0}a\varepsilon_{\mathrm{F}}%
\sqrt{2K_{2}})$ with $y_{0}=\sqrt{2A/K_{2}}$ and $\bar{t},\bar{v}_{d}$ are
dimensionless time and velocity with real time $t$, and the drift velocity,
$v_{d}$. The formation energy, $E_{\mathrm{SF}}$, and kinetic energy,
$E_{\mathrm{SK}}$, associated with a single soliton are respectively given as
follows,%
\begin{align}
E_{\mathrm{SF}}  &  =\frac{4}{3}\varepsilon_{\mathrm{F}}\sqrt{A}\frac
{K_{2}^{3/2}}{K_{4}},\label{SolitonEne}\\
E_{\mathrm{SK}}  &  =\frac{2}{3}\varepsilon_{\mathrm{F}}\sqrt{A}\frac
{K_{2}^{3/2}}{K_{4}}\bar{v}_{d}^{2}=\frac{1}{2}E_{\mathrm{SF}}\bar{v}_{d}^{2}.
\end{align}
In terms of eq.~(\ref{SolitonEne}), the number density of solitons per unit
length $n_{s}$ is given by{%
\begin{equation}
n_{s}={\frac{a}{\Delta}}\sqrt{\frac{2\pi m_{\text{S}}}{h^{2}\beta}}e^{-\beta
E_{\mathrm{SF}}}. \label{N0}%
\end{equation}
Here we introduced the rest mass of a soliton
\begin{equation}
m_{\text{S}}=E_{\mathrm{SF}}={\frac{4}{3}}\varepsilon_{\mathrm{F}}\sqrt
{A}{\frac{K_{2}^{3/2}}{K_{4}}}.
\end{equation}
Each soliton has a finite width $\Delta=2\sqrt{A/K_{2}}a$. Because of this
width, we need to take account of the excluded volume effect which leads to a
reduction factor $a/\Delta$ in }eq.~({\ref{N0}).}

Regarding the anomalous dielectric properties, we first note that dielectric
function $\varepsilon(\omega)$ is written as $\varepsilon(\omega)=1+4\pi
\sigma(\omega)/i\omega$ where $\sigma(\omega)$ is the dynamical conductivity
computed by using the Kubo formula,%
\begin{align}
\varepsilon(\omega)  &  =1+\frac{4\pi\sigma(\omega)}{i\omega}\nonumber\\
&  =1+2\pi\left(  \frac{e}{\pi}\right)  ^{2}L\left.  \mathcal{D}%
(0,0;i\omega_{n})\right\vert _{i\omega_{n}\rightarrow\omega+i\delta},
\end{align}
The thermal Green function is defined by $\mathcal{D}(q,q^{\prime};i\omega
_{n})=\int_{-\beta}^{\beta}d\tau e^{i\omega_{n}\tau}\langle T_{\tau}%
\varphi_{q}(\tau)\varphi_{-q^{\prime}}\rangle,$ with $\omega_{n}=2\pi
nk_{B}T,$ $k_{B}$ and $T$ being the Boltzmann constant and temperature,
respectively.\cite{impurity} The Fourier component $\varphi_{q}$ is introduced
by $\varphi(x)=L^{-1/2}\sum_{q}e^{iqx}\varphi_{q}$, and $\varphi_{q}%
(\tau)=e^{\tau\mathcal{H}}\varphi_{q}e^{-\tau\mathcal{H}}$ with $L=Na$ being
the system size.

According to Ref.~\cite{Krum}, the dynamical correlation function, and then
the frequency dependences of the dielectric function due to domain wall
motions is given in the dilute limit of soliton (domain wall) density as
follows, ($2t_{\text{s}}$ in Ref.~\cite{Krum} is written as $\tau_{\text{s}}$
here)
\begin{equation}
\varepsilon(\omega)\sim\varphi_{0}^{2}\frac{\tau_{\text{s}}}{1+\omega^{2}%
\tau_{\text{s}}^{2}}, \label{epsilon2}%
\end{equation}
with $\tau_{\text{s}}$ being given by $\tau_{\text{s}}^{-1}=\int dv_{d}%
\ v_{d}n_{\text{S}}(v_{d})$. The Maxwell distribution for the dilute soliton
gas is given by%
\begin{equation}
n_{\text{S}}(v)=\frac{m_{S}}{h}\frac{a}{\Delta}\exp\left[  -\beta\left(
\frac{1}{2}m_{\text{S}}\bar{v}_{d}^{2}+E_{\text{SF}}\right)  \right]  ,
\end{equation}
which gives
\begin{align}
E_{4}  &  =\frac{h}{\tau_{\text{s}}}=\frac{1}{2\beta}\frac{a}{\Delta}e^{-\beta
E_{\text{SF}}}=\frac{k_{B}T}{4}\sqrt{\dfrac{K_{2}}{A}}e^{-\beta E_{\text{SF}}%
}\nonumber\\
&  =\varepsilon_{\mathrm{F}}\left(  \frac{k_{B}T}{4\varepsilon_{\mathrm{F}}%
}\right)  \sqrt{\dfrac{K_{2}}{A}}\exp\left(  -\dfrac{4}{3}\dfrac
{\varepsilon_{\mathrm{F}}}{k_{B}T}\sqrt{A}\dfrac{K_{2}^{3/2}}{K_{4}}\right)  .
\end{align}


In. Fig.\thinspace2, we show how the characteristic frequency scale,
$\tau_{\text{s}}^{-1}$, varies as a function of $k_{B}T/\varepsilon
_{\mathrm{F}}$, with our fixed choice of $g_{D}=3$ and $g_{C}=1$. It is
clearly seen that $\tau_{\text{s}}^{-1}$ changes over 10 orders of magnitude
over the temperature window around $T\sim100$K and the range of KHz is easily obtained.

\begin{figure}[h]
\begin{center}
\includegraphics[width=8cm]{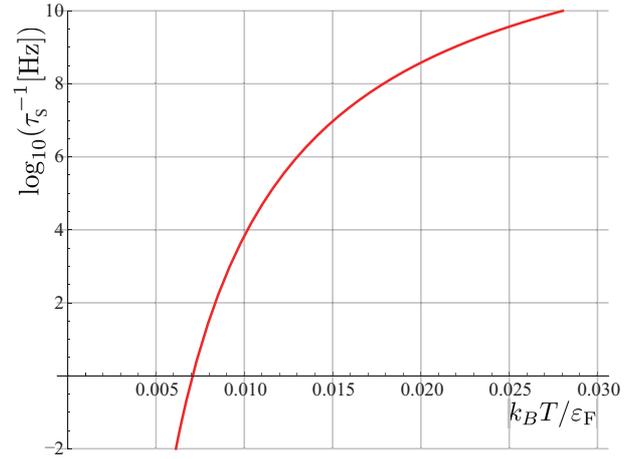} \vskip 0.5truecm
\end{center}
\caption{Characteristic frequency of {dielectric responses due to the central
modes of domain walls (solitons), as a function of }$k_{B}T/\varepsilon
_{\text{F}}$, with our fixed choice of parameters, $g_{D}=3$, $g_{C}=1$, and
$A=0.1.$}%
\end{figure}

Experimentally, however, the maximum indicate much weaker $\omega$
dependences\cite{AbdelJawad} suggesting an interesting fact that solitons may
not be moving freely and/or independently, due to mutual interactions between
solitons\cite{kishineLame} as described by the finite dispersions of the
phason mode of soliton lattice or due to impurity pinning.\cite{impurity}
These will be subjects of further studies.

In summary we have demonstrated that there are four different energy scales in
charge excitations in the Hamiltonian eq.~(\ref{PhaseH2}), as schematically
shown in Fig. 3.  
Here we point out that various effects associated with impurity or
three-dimensionality may be manifested at energies below $E_{2}$, which have
not been focused on explicitly in the present study.

\begin{figure}[h]
\begin{center}
\includegraphics[width=7cm]{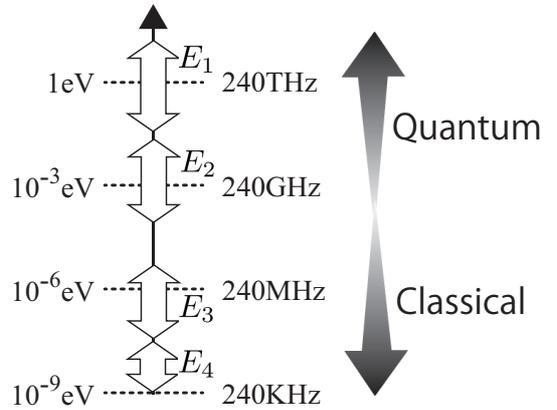} \vskip0.5truecm
\end{center}
\caption{Four different energy scales in charge excitations demonstrated in
this letter. $E_{1}$ and $E_{2}$ are in high energy quantum region, while
$E_{3}$ and $E_{4}$ are in low energy classical region. }%
\end{figure}

The present demonstrations will shed light on the possible microscopic
understanding of experimental observations\cite{AbdelJawad,Iguchi} of
dielectric anomalies seen in the range of 10 KHz based on the familiar
microscopic model Hamiltonian for electrons in solids describing the
competitions between dimer-Mott and charge ordering which are typical states
in the presence of strong correlations in quarter-filled bands. It is to be
noted that interesting phenomena observed in specific
heat\cite{Bljakovic2009,Bjarkov} and infrared optical measurements\cite{Tomic}
are in different energy region from the present studies, though these will be
mutually interrelated.

{\ Finally, we give brief comments on the relevance of the present scheme to
experimentally observed KHz dynamics in a mono-axial chiral
helimagnet\cite{KishineOvchinnikov,Togawa} near the boundary of magnetic phase
transitions.\cite{Mito} This will be studied in detail elsewhere.}

\begin{acknowledgment}
{
Special thanks are due to T. Sasaki for very informative discussions in various stages. H.F. thanks H. Kishida, H. Seo and H. Yoshioka for discussions in early stage of the work. We thank M. Mito and A. S. Ovchinnikov for fruitful discussions.
This work was supported by  a Grant-in-Aid for Scientifc Research (A) (No.\ 15H02108),  (B) (No. 17H02923) and (S) (No. 25220803) from the MEXT of the Japanese Government, and the JSPS Core-to-Core Program, A. Advanced Research Networks. }
\end{acknowledgment}

\end{document}